\documentstyle[epsfig,graphics]{elsart}

\def\apj{{\it ApJ}}
\def\araa{{\it ARA\&A}}

\def\mnras{{\it MNRAS}}

\def\la{\hbox{{\lower -2.5pt\hbox{$<$}}\hskip -8pt\raise
-2.5pt\hbox{$\sim$}}}
\def\ga{\hbox{{\lower -2.5pt\hbox{$>$}}\hskip -8pt\raise
-2.5pt\hbox{$\sim$}}}

\begin{document}
\begin{frontmatter}
\title{The GZK Feature in our Neighborhood of the Universe}
\author[ad1]{Michael Blanton},
\author[ad1]{Pasquale Blasi}
\author[ad2]{Angela V. Olinto\thanksref{corr}},
\thanks[corr]{Corresponding author. E-mail: olinto@oddjob.uchicago.edu}
\address[ad1]{Fermi National Accelerator Laboratory, Batavia, IL
60510-0500}
\address[ad2]{Department of Astronomy \& Astrophysics,  \& Enrico Fermi
Institute, \\ The University of Chicago, Chicago, IL 60637}

\begin{abstract}

We calculate numerically the spectrum of ultra-high energy
cosmic rays on Earth assuming that their sources are distributed in
space like the observed galaxies.  We use the CfA2 and 
the PSCz galaxy redshift surveys to model the local galaxy
distribution, properly taking into account the galaxy
selection functions for each survey. When the survey selection effects
are included, we find that  the local overdensity is only a
factor of two, an order of magnitude less than used in some earlier
studies. An overdensity of two is not enough to bridge the gap between 
the  predicted number of events above $10^{20}$ eV and the
measured flux at these highest energies.  This conclusion is particularly
strong for soft injection spectra ($\propto E^{-3}$)  where the observed
number of events is 6 $\sigma$ higher than the expected one. However, if
the injection spectrum is hard ($\propto E^{-2}$), the small local
overdensity helps bring the present data within 2$\sigma$ of the low
number  of events predicted above $10^{20}$ eV. In this case, the
Greisen-Zatzepin-Kuzmin cutoff is not  a {\it cutoff} but rather a {\it
feature} in the cosmic ray spectrum.

\end{abstract}

\end{frontmatter}

\section{Introduction}
The unexpected detection of  cosmic rays with
energies above $10^{20}$ eV  has triggered considerable interest
in the  possible origin and nature of these particles \cite{data}.
These highest energy events are surprising for the following reasons.
If these particles are protons, they  likely originate in extragalactic
sources, since  at these high energies the Galactic magnetic field cannot
confine protons in the Galaxy. However, extragalactic protons with
energies above  a few times $10^{19}$ eV can produce pions through
interactions with the cosmic microwave  background (CMB) and consequently
lose significant amounts of energy as they traverse intergalactic
distances \cite{GZK66}.  Therefore, in addition to the extraordinary
energy requirements for astrophysical sources to accelerate protons to
$\ga 10^{20}$ eV, the photopion threshold reaction  suppresses the
observable flux above $\sim 10^{20}$ eV. These conditions were expected
to cause a natural high-energy limit to the cosmic ray spectrum known as
the Greisen-Zatsepin-Kuzmin (GZK) cutoff \cite{GZK66}.

As clearly shown by the most recent compilation of the AGASA
collaboration data \cite{haya00}, the spectrum of cosmic rays does not end
at the expected GZK cutoff. The significant flux observed above  $10^{20}$
eV together with a nearly isotropic distribution of event arrival
directions has challenged astrophysically based scenarios (see e.g.,
\cite{Olin00} and references therein) and has inspired a number of 
alternatives (see, e.g., \cite{BS00} and references therein). However, 
the GZK cutoff is not an absolute {\it end} to the cosmic ray spectrum
but it generates a {\it  feature} around $5 \times 10^{19}$ eV.  The
spectrum recovers at energies above this feature \cite{BG88} and the local
distribution of sources can significantly  affect the agreement between
predicted and observed spectra. In particular, a
local overdensity of sources can decrease the gap between
observed and detected  events above the GZK cutoff 
\cite{BG79,MT99,BW00}. This effect can easily be understood: 
photopion energy losses limit the maximum distance at which sources can
contribute at the highest energies to a few tens of Mpc,
while cosmic rays below the pion
production threshold come from much larger volumes.  A local overdensity
will increase the observed flux at the highest energies relative to
the lower energy flux.  Here we show that the observed local overdensity is
not high enough to explain the data unless, perhaps, the sources have a hard
injection spectrum.  

In the following, we simulate the propagation of ultra-high energy 
cosmic rays (UHECRs) from extragalactic sources to Earth.  Our numerical
propagation code includes pair production, photopion production, and
adiabatic losses due to the universe's expansion.  For a uniform
distribution of sources, our numerical results agree well with previous
results and  analytical calculations. We study the
effect of a local overdensity by assuming that the number density
of sources is proportional to the number density of galaxies. To keep this
study independent of specific hypothetical sources, we use different
injection spectra and consider a possible source evolution with
redshift. However, since we assume that UHECRs are protons, our results
are mostly relevant for extragalactic astrophysical acceleration
scenarios. 

Rather than adopting analytic models of the galaxy distribution, as do 
\cite{BW00,Ptu98}, we extract the distribution of galaxies from
observations of large scale structure using the CfA2 and the PSCz galaxy
catalogs. We assume that the density field derived from these studies has
the same shape as the density field of UHECR sources, although {\it a
priori} UHECR  sources may cluster differently from luminous matter.  We
show that the local density is only about a factor of two above
the mean, in contrast with much higher estimates previously published 
\cite{MT99}. This large discrepancy was caused by neglecting
the necessary galaxy selection functions which account for the fact that
nearby galaxies are far easier to detect than far away galaxies. Once
we include the selection functions, we find that the real overdensity is
not high enough to bridge the gap between predicted and observed spectra
for soft sources ($J(E) \propto E^{-\gamma}$, with $\gamma = 3$). However,
sources with a  hard injection spectrum ($\gamma
\la 2.1$) can fit the present data within 2$\sigma$, at energies above
a few $10^{19}$ eV. Sources of UHECRs distributed as ordinary galaxies 
are marginally consistent with present data and,
for hard injection spectra, the GZK cutoff is not really a {\it cutoff}
but a {\it feature} in the high-energy cosmic ray spectrum.

The plan of this paper is as follows. In \S2, we discuss the proper
way to model a distribution of sources associated with
galaxies. We derive the galaxy density field in our neighborhood of the
universe using the CfA2 and PSCz galaxy redshift surveys with their
respective selection function.  In \S3, we describe our UHECR
propagation code. In \S4, we display the results for different injection
spectra and a realistic spatial distribution of sources. We also contrast 
our results with previous work and  analytical estimates. We conclude in
\S5.

\section{The Galaxy Density Field}

Although galaxies are not homogeneously distributed in the 
local universe, the GZK cutoff is usually calculated assuming a  
homogeneous distribution of sources throughout space. If the sources are 
distributed like the luminous matter around us, the effects of an 
inhomogeneous galaxy distribution needs to be taken into account when 
predicting the spectrum of UHECRs.  In order to include the effects of 
the spatial inhomogeneity in the UHECR spectrum, we must consider the 
estimated galaxy density field in our neighborhood of the universe.   The 
density field is usually measured by selecting galaxies from an imaging 
survey of the sky and taking their redshifts. Almost invariably, the 
galaxies are selected to be brighter than some limiting flux in some band, 
$f_{\mathrm{lim}}$,  expressed as an ``apparent 
magnitude,'' $m_{\mathrm{lim}}=-2.5\log_{10} (f_{\mathrm{lim}}/f_0)$, 
where $f_0$ is an arbitrary zero-point. For all (or for some random 
subsample) of the galaxies brighter than this, one takes their spectra 
and determines their redshifts $z$. According to the Hubble law, the 
redshifts are related to their distances $ d = H_0 c z$, 
where $c$ is the speed of light and $H_0 
\equiv 100 h$  Mpc/ km/s, with $h  \approx 0.5$--$0.8$ ({\it e.g.}, 
\cite{branch98a}).  
 
However, a flux-limited survey is not  a volume-limited survey. In a 
flux-limited survey, the raw distribution of redshifts cannot be used 
without regard to the way galaxies were selected. Here we describe 
the proper way to derive density fields from galaxy redshift surveys. We 
limit ourselves to measuring the density in redshift space and do not 
include the small effects of deviations from the Hubble law due to 
galaxy peculiar velocities. We base our approach on methods dating back 
to \cite{davis82a} (see, e.g.,  \cite{strauss95a} for a recent review). 
 
We will use two surveys for our analysis. 
To compare directly with \cite{MT99}, we consider the Center 
for Astrophysics Redshift Survey (CfA2; \cite{huchra95a}).
Although this 
survey comprises about 10,000 galaxy redshifts (selected to be brighter 
than $m = 15.5$, a $B$-band magnitude), it  
covers only about 17\% of the whole sky. In order to better 
evaluate the effects of the density field of galaxies on the cosmic ray 
spectrum, we should use surveys which probe the density field over nearly 
the whole sky. The best sample of galaxies to use for this purpose is the 
IRAS PSC$z$ Survey \cite{saunders00a}, which consists of about 
15,000 galaxies with infrared fluxes $>0.6$ Jy and covers about 84\% of 
the sky. 
 
A consequence of the flux limits in any survey is that at different 
redshifts, a different set of galaxy luminosities $L$ is observed, 
determined by the faintest luminosity observable at that redshift 
$L_{\mathrm{min}}(z)$. For an Euclidean metric, this luminosity is 
related to the flux limit by: 
\begin{equation} 
L_{\mathrm{min}}(z) = 4 \pi (H_0 c z)^2 f_{\mathrm{lim}}. 
\end{equation} 
At cosmological distances (generally only appropriate when $z>0.1$) 
more complicated relations apply, which in general depend on the 
cosmological model \cite{peebles93a} (see \cite{hogg99a} for a 
useful compilation of results).  If the distribution of galaxy 
luminosities is described by the galaxy luminosity function $\Phi(L)$, 
then the  fraction of all galaxies  that are 
observable at any redshift is given by: 
\begin{equation} 
\phi(z) = \frac{\int_{L_{\mathrm{min}}(z)}^\infty dL \ \Phi(L)} 
{\int_{0}^\infty dL \  \Phi(L)}. 
\end{equation} 
The quantity $\phi(z)$ is usually referred to in the literature as the 
``selection function'' \cite{peebles80a}. The most common methods 
used to determine the galaxy luminosity function from the survey 
itself are those of \cite{efstathiou88a} and 
\cite{sandage79a}. These methods assume that the luminosity function 
is universal (i.e., independent of redshift) and use nearby galaxies to 
determine the shape of the faint end and far away galaxies for 
the shape of the bright end. 
 
Consider, for example, Figure 1, which shows in the top panel 
the distribution of galaxies and redshifts in CfA2.  Here we express 
galaxy luminosity in terms of the ``absolute magnitude,'' $M = 
-2.5 \log_{10} L +$ const; thus, in the figure, the faintest galaxies 
are at the top.  The thick solid line shows the flux limit of the 
survey, translated into an absolute magnitude limit at each 
redshift. Because of this limit a number of galaxies which are 
observable at low redshifts are too faint to be observed at higher 
redshift. The fraction of galaxies $\phi(z)$ between absolute 
magnitudes $-22<M<-10$ which are unobservable at each redshift is 
shown in the bottom panel of Figure 1, based on a fit to the luminosity 
function in the survey using the method of \cite{efstathiou88a}. Because 
the function falls rapidly from unity, it is clear that even at low 
redshifts the effects of the flux limit are important.  

\begin{figure}[thb] 
 \begin{center} 
  \mbox{\epsfig{file=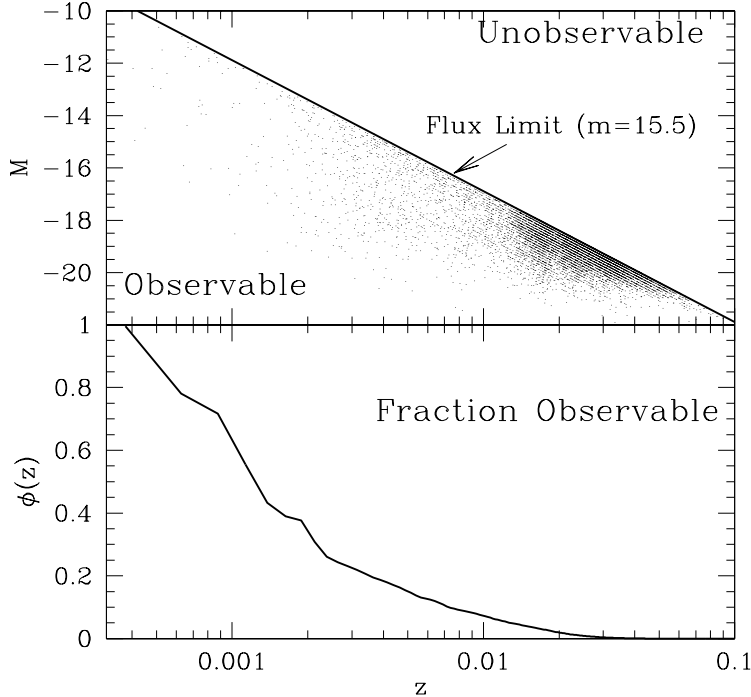,width=11cm}} 
  \caption{\em {The top panel shows the absolute magnitudes 
(related to luminosity by $M = -2.5 \log_{10} L +$ const) and 
redshifts of CfA2 galaxies. Shown as the thick solid line is the 
flux limit, converted to the appropriate absolute magnitude at each 
redshift. The bottom panel shows the fraction of galaxies in the range 
$-22<M<-10$ that we estimate to be brighter than the flux limit. This 
function falls rapidly with redshift. When interpreting the top plot, 
remember that the volume probed at low redshift is far smaller than 
that probed at high redshift. 
}} 
 \end{center} 
\end{figure} 
 
In principle, we can construct a 
volume-limited sample by choosing only galaxies brighter than some 
magnitude $M$ up to the  redshift at which the 
thick solid line representing the flux limit crosses $M$. This usually 
decreases significantly the number of galaxies and  the redshift
depth of the survey. A more effective approach is to  use 
$\phi(z)$ to calculate the distribution of observed galaxies 
with redshift which one would expect if the distribution were homogeneous.
Then, the density field can be inferred by comparing the actual counts to
these expected counts. The  top panel of Figure 2 compares these expected
counts (dotted line) in  redshift shells of thickness 0.001 to the
observed counts (solid line) in  CfA2. It appears that locally we are in
an overdensity of galaxies of  about a factor of two; note that at large
distances, where each shell  corresponds to a considerable amount of
volume, and thus averages over  large-scale structures, the number of
galaxies is very nearly  the expected number. Instead, if  the flux limit
is neglected (dashed  line), the ``expected'' number of galaxies in each
shell scales  approximately as the square of the redshift of that shell.
If we normalize these ``expected'' counts in
approximately the same way as \cite{MT99}, we recover the incorrect 
result that we live in an overdensity of approximately a factor of 30. It 
is clear from the comparison of the dotted to the dashed curve that 
neglecting the flux limits is a poor approximation. Put simply, we do not 
live in a large overdensity, but we can certainly detect galaxies more 
easily if they are close by rather than far away.  
\begin{figure}[thb] 
 \begin{center} 
  \mbox{\epsfig{file=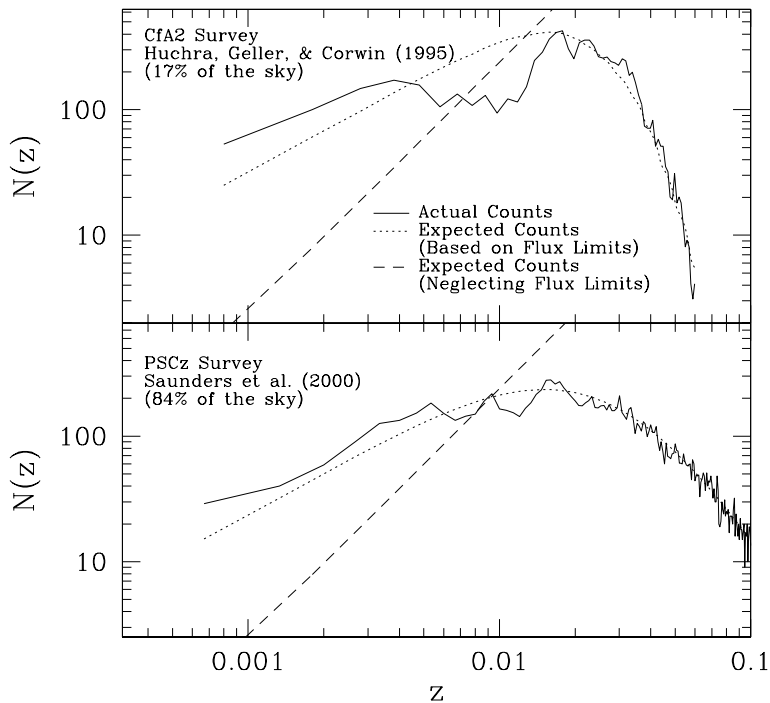,width=11cm}} 
  \caption{\em {Comparison of observed counts ({\it solid 
line}) to those predicted based on the flux limits ({\it dotted line}) 
and those predicted neglecting the flux limits ({\it dashed 
line}). The CfA2 survey is shown at top, the PSC$z$ at bottom. Both 
show a local overdensity of only about a factor of two when the flux 
limits are properly accounted for. 
}} 
 \end{center} 
\end{figure} 

As mentioned above, the CfA2 survey covers a relatively small fraction 
of the sky. Thus, the PSC$z$ redshift survey provides a more useful 
sample to use in the context of this paper. Using the selection 
function provided by \cite{saunders00a}, we again show the expected 
versus the observed counts for the PSC$z$ survey in the bottom panel 
of Figure 2. This survey also shows we are living in a 
slight overdensity, and furthermore reveals the general homogeneity of 
the nearby universe. (The actual counts and their dependence on 
redshift are  different from CfA2, because the galaxies 
are selected in different ways). 
 
\section{The Propagation Code for UHECRs} 
 
Armed with a more realistic model of the local universe, we can calculate 
the spectrum of UHECRs that would be observed on Earth for  
extragalactic sources with different injection spectra. 
Our numerical propagation code includes pair production and photopion 
production as energy losses, as well as adiabatic losses  
due to the expansion of the universe. We compare our  numerical   results 
with   analytical calculations which we carry out as in \cite{BG88}. In 
order to isolate the effect of density inhomogeneities  we neglect the 
effect of magnetic fields in this paper.
   
We compare the results of the simulations with  the  
observed spectrum by requiring that the total number of simulated events 
with energy above a normalizing energy, $E_{norm}$, be the same as 
what is observed.  In general we choose $E_{norm} \simeq 10^{19}$ 
eV  since  at lower energies the flux of cosmic rays is likely to have a
galactic origin and at higher energies the  observed number of events is
very small. For a given source spectral index, we generate many
realizations of the spectrum on Earth. Each  realization has the same
total number of observed particles above  $E_{norm}$ calculated as
follows. The   flux of UHECRs contributed from a shell of thickness  
$\Delta z$ at redshift $z$ is proportional to $p(z) \propto (1+z)^{m-5/2}
f(z)$, where the redshift dependence of the density of 
sources and the flux suppression due to distance are 
included explicitly \cite{BBDGP90}. In this formula, we allow for
source luminosity  evolution through the parameter $m$. The function 
$f(z)$ describes possible deviations  from a homogeneous spatial 
distribution of  sources. For a homogeneous distribution $f(z)=1$, 
otherwise $f(z)$ represents the local overdensity of sources at redshift 
$z$ as, for example, those derived from the catalogs introduced in \S2. 
We assign the redshift of one event by extracting a random number  
according to the distribution $p(z)$ given above. The energy of the
particle at  the  source is extracted from a distribution representing the
spectrum of  the source assumed to have the form of a power
law $E^{-\gamma}$.  The particle is then propagated from the source to
the detector. 
 
The photopion energy losses are simulated following 
\cite{Acht98}. In each spatial step  of the simulation 
of size $\Delta s$ ($\sim 200$ kpc),
the average number of photons that can induce the production 
of pions in the scattering with a proton of energy $E$ at time 
$t$ is 
\begin{equation} 
\langle N_{ph}(E,\Delta s)\rangle = \frac{\Delta s}{K_p(E) l(E)}.
\end{equation}
Here, $l(E)=c\left[(1/E)(dE/dt)\right]^{-1}$ is taken from \cite{BG88} and
$K_p(E)$ is an effective inelasticity at energy $E$, which can be
approximated by
\begin{equation}
K_p(E)\simeq 0.2\left(\frac{E_{th}+2.5E}{E_{th}+E}\right), 
\end{equation}
with $E_{th}=3\times 10^{11}$ GeV \cite{Acht98}. Once the 
average number of pion producing photons has been determined over the 
path $\Delta s$, the actual number of photons with which the proton interacts
is extracted from a Poisson distribution with mean $\langle
N_{ph}(E,\Delta s)\rangle$.  For sufficiently small $\Delta s$, the number
of photons in each step is usually either zero or one. The energy of each
photon is extracted from a Planck distribution at temperature
$T_{CMB}=2.728$ K (with a minimum energy corresponding to the threshold for
pion production). Since the CMB photons are isotropically distributed in
space, we extract the interaction angle from  a flat distribution in
$\cos(\theta)$. The final energy of the proton after each scattering
with a CMB photon is calculated from the kinematics of the scattering. 
Since the inelasticity for pair production is very low, we treat it as
a continuous energy loss process. 

We compare our numerical results with analytic results 
for the modification factor from single sources and from a 
distribution of sources as in \cite{BG88}. The agreement is excellent  and
the effect of the fluctuations  at energies larger than
$\sim (3-4)\times 10^{19}$ eV is evident in Figures 5--8 as we discuss
below. The average of the simulated flux is slightly larger than the
analytic one, as expected for the stochastic 
process of photopion production (on small distances there is an
appreciable chance that some protons do not interact at all).

\section{Results}

As we stressed above, a local overdensity increases the observed flux
above GZK energies with respect to the flux at lower energies. To show
this effect clearly, we estimated the change in the spectrum due to a
simple top-hat model  before considering more realistic models of the
galaxy density field. In Figure 3, we display the results of our analytical
calculation for three choices of the overdensity,
$\rho/\bar\rho=1,~10,~30$ 
(solid, dashed, and dash-dotted lines respectively),
all in a volume of radius 
$\sim 20$ Mpc around the Earth and with source spectral index $\gamma=3$.
From this figure, it is clear that the overdensity increases the flux at
the highest energies versus the flux at and below the GZK feature, as
mentioned above. 

\begin{figure}[thb]
 \begin{center}
  \mbox{\epsfig{file=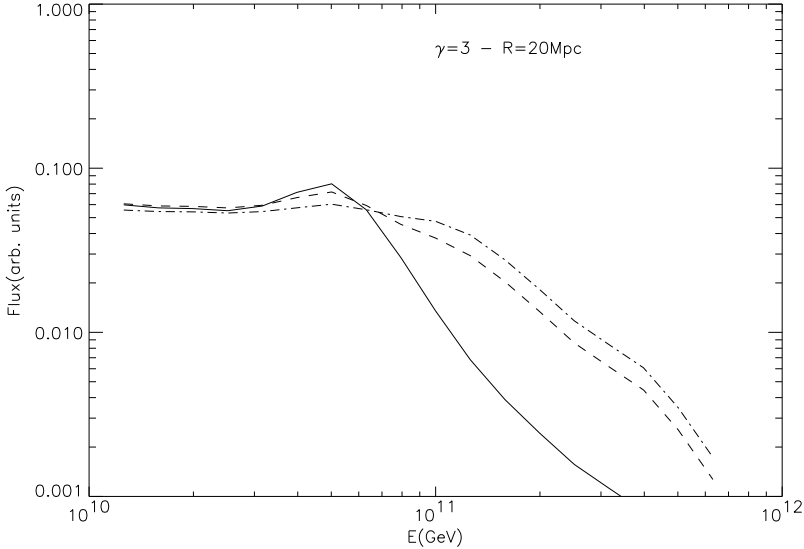,width=11cm}}
  \caption{\em {The effect of a local overdensity within $20$ Mpc on the
fluxes of UHECRs, according to our analytical
calculation for three choices of the overdensity,
$\rho/\bar\rho=1,~10,~30$ 
(solid, dashed and dash-dotted lines respectively),
all in a volume of radius 
$\sim 20$ Mpc around the Earth and with source spectral index $\gamma=3$.
}}
 \end{center}
\end{figure}

The galaxy surveys discussed in \S2 provide a more realistic model of the
local density field. Among the different surveys, the PSCz catalog covers 
a lot more solid angle and reaches further (up to a  redshift  
$z_{max}=0.1$) when compared to the CfA2 survey. Thus, we use the PSCz to 
study the effect of different source spectra below. But first we compare 
the results of the two catalogs for a fixed spectral index ($\gamma = 3$) 
with the homogeneous distribution in Figure 4.  This choice of 
$\gamma$  allows a direct comparison of our results with those of 
\cite{MT99}. In this figure, we normalize the  
spectrum by requiring that the number of events with energy above  
$10^{19}$ eV  equal the AGASA number of 728 \cite{AGASA,AGASAnew}.  
The error bars  in the simulation are obtained by generating 100
realizations and  calculating the mean and variance of the set.  Figure
4 shows that the two catalogs give very similar results.

\begin{figure}[thb]
 \begin{center}
  \mbox{\epsfig{file=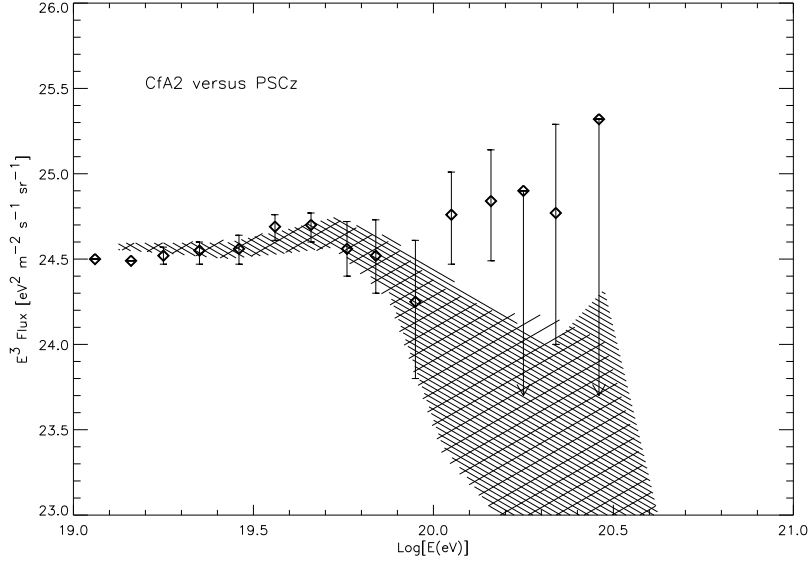,width=11cm}}
  \caption{\em {Simulated fluxes of UHECRs for $\gamma=3$ using the
CfA2 catalog (hatched upwards to the right, i.e., like ``/'') and the PSCz
catalog (hatched downwards to the right, i.e., like ``$\setminus$''). }}
 \end{center}
\end{figure}

In Figure 5, we compare the case of a homogeneous distribution and the
PSCz catalog versus the AGASA data also for $\gamma=3$.  
The total number of events with $E>10^{20}$ 
eV is $1.2\pm 1.0$ for the homogeneous case and $1.5 \pm 1.0$ for the 
PSCz catalog of galaxies. Both numbers are consistent with 1, while 
AGASA has detected 8 events with $E>10^{20}$ eV. The data is more than 
6 $\sigma$ away from the observations. None of our realizations have 
the observed number of events for this soft spectrum. 
In the figure, the solid and dashed lines 
represent the result of the analytical calculations for 
the same value of the parameters and for the homogeneous and PSCz cases 
respectively. The dash-dotted and dash-dot-dot-dotted lines trace 
the mean simulated fluxes for the homogeneous case (hatched downwards to
the right, i.e., like ``$\setminus$'') and  the PSCz case (hatched
upwards to the right, i.e., like ``/''). As can be seen from Figure 5,
the difference between the analytic  calculations and the mean of the
numerical calculation is quite small.   In addition, the homogeneous case
has only slightly lower fluxes at higher  energies than the realistic
models. Here  our results differ   significantly from those of
\cite{MT99}, due to the fact that the real local overdensity is much
smaller than  found in  \cite{MT99}.

\begin{figure}[thb]
 \begin{center}
  \mbox{\epsfig{file=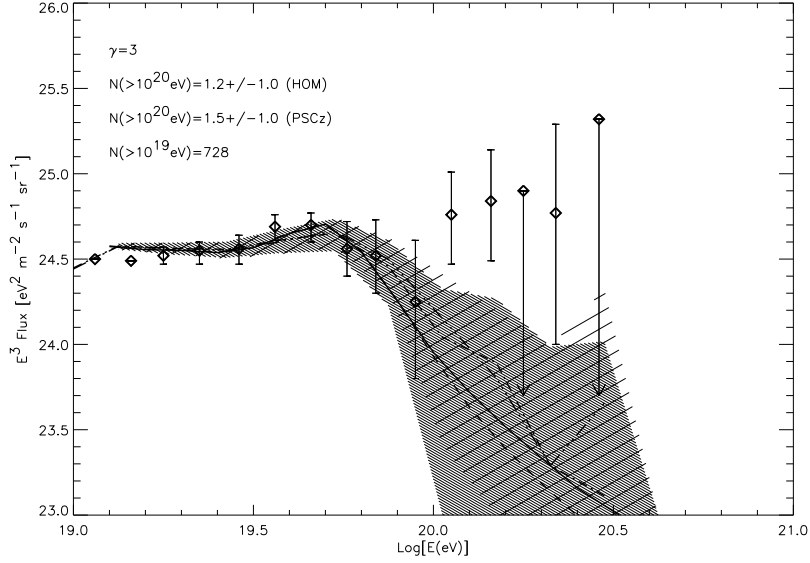,width=11cm}}
  \caption{\em {Simulated fluxes for the AGASA statistics of 728 events above 
$10^{19}$ eV, and $\gamma=3$, using a homogeneous source 
distribution ($\setminus$ hatches)
and the PSCz distribution (/ hatches). The solid and dashed lines
are the results of the analytical calculations for the same two cases. 
The dash-dotted and dash-dot-dot-dotted lines trace 
the mean simulated fluxes for the homogeneous and the PSCz cases. 
}}
 \end{center}
\end{figure}
 
In Figure 6, we show the generated fluxes assuming that 9075 
events have been observed above $10^{19}$ eV, a number which the Auger 
project is expected to reach in the first 3 years of operation. Again we 
compare the results of a homogeneous distribution of sources
($\setminus$ hatches) with those  associated with the PSCz galaxy
distribution (/ hatches),  using $\gamma = 3$ for both. As expected from
a larger sample,  the size of the error bars decreases significantly. The
number of  events at
$E>10^{20}$ eV  in the two cases is still much  smaller than  if
extrapolated from the AGASA observations ($\sim 100$). This figure shows 
that the Auger project will be able to constrain the nature of the GZK
feature much more accurately than possible at present.
\begin{figure}[thb]
 \begin{center}
  \mbox{\epsfig{file=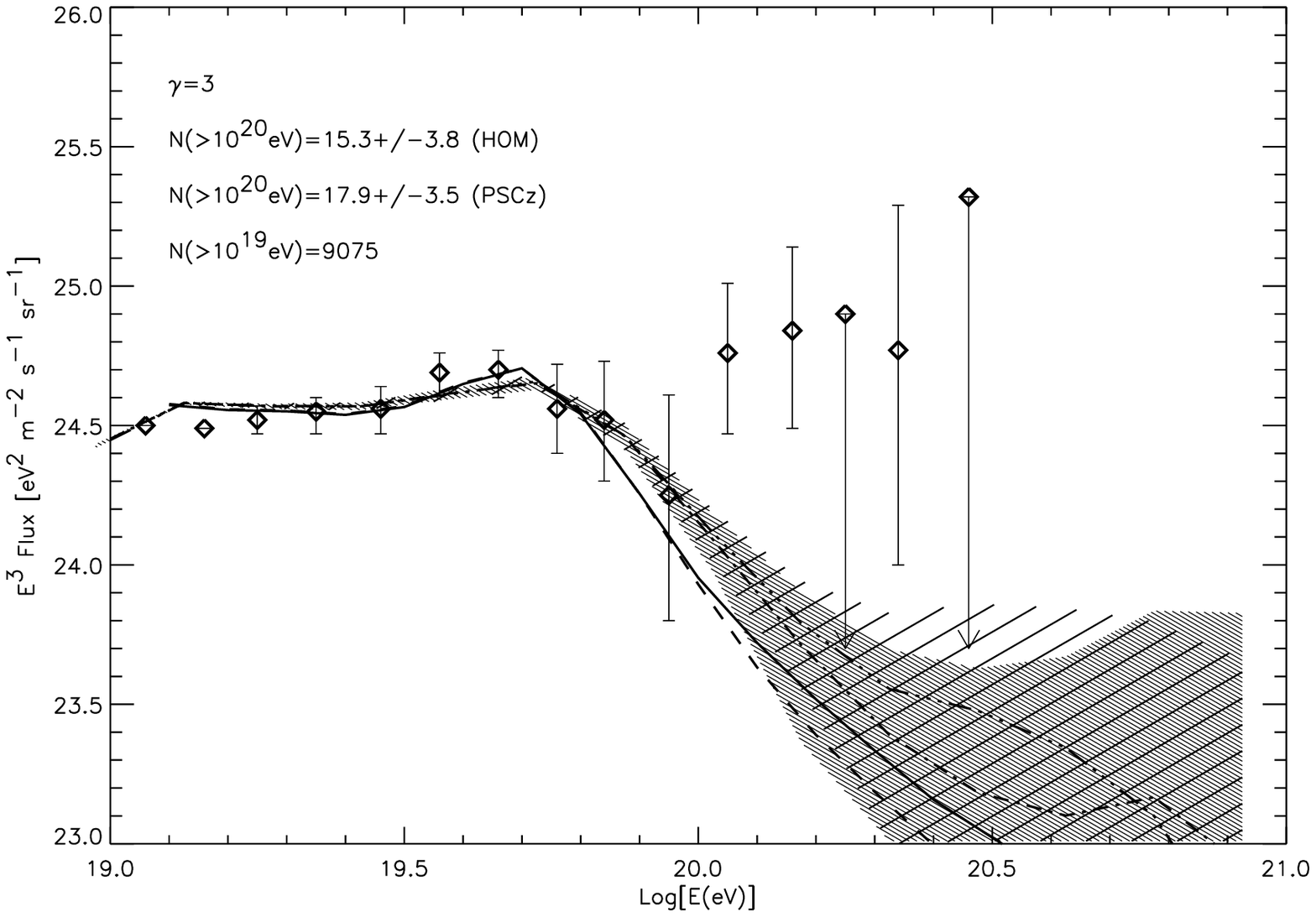,width=11cm}}
  \caption{\em {Simulated fluxes for the Auger projected 
statistics of 9075 events above 
$10^{19}$ eV, and $\gamma=3$, using a homogeneous source 
distribution ($\setminus$ hatches)
and the PSCz distribution (/ hatches). The solid and dashed
lines are the results of the analytical calculations for the same two
cases.  The dash-dotted and dash-dot-dot-dotted lines trace 
the mean simulated fluxes for the homogeneous and the PSCz cases. 
}}
 \end{center}
\end{figure}
 
It is clear that with a source injection spectrum as steep as $\gamma = 3$, 
the number of events with $E>10^{20}$ eV is significantly smaller than
what  has been observed so far. Since sources with harder spectra have
been  proposed, we consider different choices for the injection
spectrum. We simulated the case of $\gamma = 2.7$, but the results do not
diffr significantly from the $\gamma = 3$ case. Next we discuss the case
of $\gamma = 2.1$, where we find the gap between observed and predicted
flux to reach the 2 $\sigma$ level.
 
In Figure 7, we show the results for $\gamma=2.1$ for three cases: a
homogeneous distribution of the sources  with $z_{max}=0.1$ (/ hatches),
the PSCz distribution with $z_{max}=0.1$ (horizontal hatches), and a
homogeneous distribution with $z_{max}=1$ ($\setminus$ hatches). For
such  hard spectra,  we need to adjust the normalization energy,
$E_{norm}$,  such that the flux  at energies below $10^{20}$ eV is
consistent with observations in this energy range.  Thus,
we choose the normalization  energy, $E_{norm}= 4\times  10^{19}$ eV,
where the number of events observed by AGASA \cite{AGASAnew}   (49) is
still statistically significant.

\begin{figure}[thb]
 \begin{center}
  \mbox{\epsfig{file=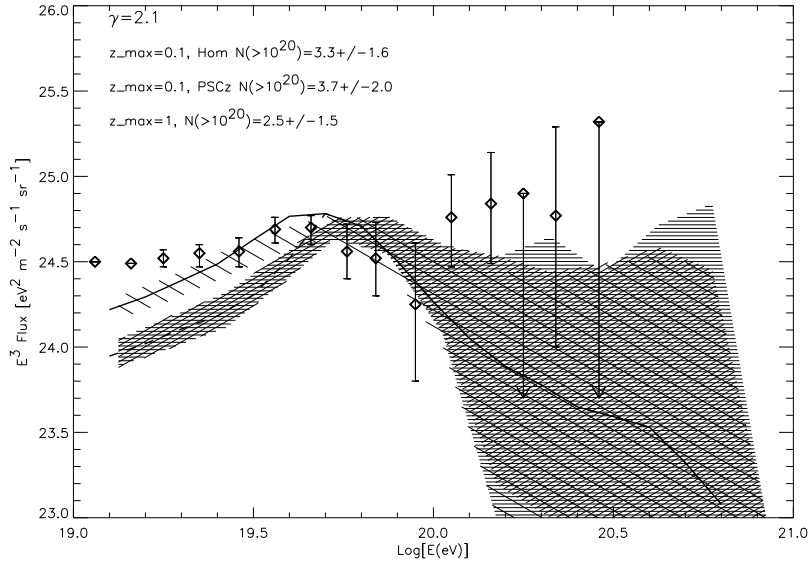,width=11cm}}
  \caption{\em {Simulated fluxes for the AGASA 
statistics of 728 events above 
$10^{19}$ eV, and $\gamma=2.1$, using a homogeneous source 
distribution with $z_{max}=0.1$ (/ hatches),
the PSCz distribution with $z_{max}=0.1$ (horizontal hatches), and
a homogeneous source 
distribution with $z_{max}=1$ ($\setminus$ hatches).
}}
 \end{center}
\end{figure}
  
Again, Figure 7 shows that the number of events at $E>10^{20}$ eV is 
affected by the local  distribution of the sources. For the adopted
normalization, the  homogeneous distribution gives 
$3.3 \pm 1.6$ events above $10^{20}$ eV (to be compared with the observed 
8) and the PSCz distribution provides $3.7\pm 2.0$ events in the same 
range.  In this last case, about $5\%$ of our realizations give a 
number of events above 
$10^{20}$ eV which is equal or larger that the observed one (consistent with  
a $2\sigma$ significance for a Gaussian error distribution).  
 
The deficit of events at energies lower than $\sim (3-4)\times 10^{19}$ eV  
for hard spectra is evident in the case of $z_{max}=0.1$. Since the 
high redshift sources 
contribute mainly to the low energy flux,  we also considered the case 
where the maximum redshift is $z_{max}=1$. The increase in $z_{max}$ moves 
the deficit to energies lower than $\sim 2\times 10^{19}$ eV, as shown in 
Figure 7 (the discrepancy is about a factor 1.6).  
 
We also considered the effect of a source luminosity evolution. For 
instance, if the luminosity of the sources increases with redshift 
$z$, the  flux of UHECRs at energies below 
the GZK cutoff will also increase. This is shown in Figure 8 for the case 
$m=0$ (horizontal hatches), $m=2$ (/ hatches), and $m=4$
($\setminus$ hatches). This type of source evolution by itself does not 
improve significantly the  agreement between the theoretical prediction 
and the AGASA data at low energies, leaving unaltered the number of events 
above $10^{20}$ eV. In fact the height of the bump at $\sim (3-4)\times 
10^{19}$ eV  also increases with large values of $m$,  thus reducing the 
number of events with energy larger than $10^{20}$ eV when the 
number of events is normalized to the integral flux  above $4\times 10^{19}$ 
eV. The number of events above $10^{20}$ eV is 
$2.8\pm 1.4$ for the case $m=2$ and $2.5\pm 1.5$ for $m=4$ 
(these numbers are for a homogeneous distribution of the sources). 
 
\begin{figure}[thb]
 \begin{center}
  \mbox{\epsfig{file=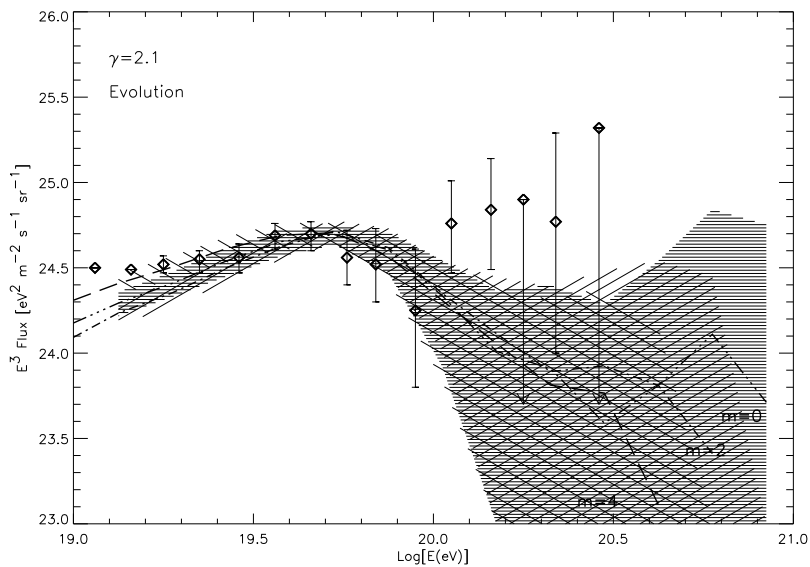,width=11cm}}
  \caption{\em {Simulated fluxes for the AGASA 
statistics of 728 events above 
$10^{19}$ eV, and $\gamma=2.1$, using a homogeneous source 
distribution with $z_{max}=1$ and $m=0$ (horizontal hatches), $m=2$ 
(/ hatches), and $m=4$ ($\setminus$ hatches). The lines are the
results of the analytical calculations, as in \cite{BG88}.
}}
 \end{center}
\end{figure}
  
In trying to bridge the gap between predicted and observed ultra-high 
energy fluxes, we should consider some additional unknowns that may 
affect the output spectrum. For instance, it is not clear how the 
extragalactic component we have considered above gets modified by a 
Galactic component at the highest energies.  The spectrum of Galactic 
cosmic rays might extend up to  $\sim 10^{19}$ eV or even to higher 
energies. If the Galactic component continues to higher energies with the 
same spectrum as the observed one at lower energies (i.e., $E^{-3.1}$),  
it will improve the agreement with the AGASA observations when combined 
with the extragalactic  flux plotted in Figure 7. We do not have 
sufficient information about the Galactic contribution at such high 
energies to rule out this  possibility. 
 
Another aspect of this problem that should be considered is the effect of 
extragalactic magnetic fields. As  shown in \cite{BO99,SLB99}, a large 
scale magnetic field with strength $\sim 10^{-7}$ -- 
$10^{-9}$ G can steepen  the spectrum of the observed 
cosmic ray particles significantly. This effect would improve the 
agreement with the AGASA data. But as the diffusive limit is reached, 
the effect might go in the opposite direction: for instance, if  the 
effective diffusion coefficient is too small, then the propagation time 
may be larger than the age of the universe for low energy particles 
generated beyond some distance, considerably reducing the  cosmic ray 
horizon \cite{stanev}. In other words a smaller portion of the universe would  
contribute to the observed flux, thus reducing it. The highest energy 
component of the spectrum should not be affected appreciably by diffusive 
propagation, since most of the particles  with $E>10^{20}$ eV  
originate in the local universe, where propagation  is likely to be 
non-diffusive. A complete answer to this question 
awaits a full  propagation code that includes the 
effects of a realistic model of extragalactic magnetic fields.

\section{Conclusion} 
 
We have studied the effects on the spectrum of ultra-high energy cosmic 
rays of sources associated with a realistic model for the galaxy 
distribution provided by the PSCz and CfA2 catalogs. We considered 
different injection spectra and the possibility of luminosity evolution 
of UHECR sources
with redshift. We showed that when the galaxy selection functions are 
properly accounted for, the local overdensity is not as large as 
found by \cite{MT99}. Thus, the AGASA observations are {\it inconsistent} with 
the predicted flux above $10^{20}$ eV for soft  injection spectra ($\gamma 
\simeq 3$).  
 
We confirm that a local overdensity helps bridge the agreement between 
theory and observations but only (and only slightly) for hard  
injection spectra ($\gamma \simeq 2$). In this case, the observations 
are within  
2$\sigma$ of the predicted number 
of events, in agreement with the findings of \cite{BW00}.   
Sources of UHECRs having a density field following that of 
galaxies are therefore consistent with present data only for hard injection 
spectra and, in this case, the GZK cutoff is not really a {\it cutoff} but 
a {\it feature} in the overall cosmic ray spectrum.
 
As future experiments such as the Pierre Auger Project \cite{Auger}, 
the High Resolution Fly's Eye \cite{HiRes}, the Telescope Array \cite{TA}, 
and the EUSO \cite{EUSO} and OWL \cite{OWL} satellite experiments increase 
the number of events observed above 
$10^{20}$ eV, a better determination of the shape of the GZK 
feature will be obtained. The GZK feature will contain information 
both on the injection spectrum (i.e., $\gamma$) as well as the 
clustering properties of the sources. Given 
$\gamma$, the clustering properties, and the angular distribution of 
arrival directions, a population of sources might be identifiable.  
If associated with active galaxies or some other specific class 
of astrophysical objects, the shape of the GZK feature can be 
further used to constrain intergalactic magnetic fields \cite{BIGM} as well 
as more exotic pieces of physics, such as violations of Lorentz  
invariance \cite{VLI}.  
 
Alternatively, the gap between observed and predicted flux may widen as
more data accumulates. This would indicate that astrophysical proton
accelerators are unlikely sources of UHECRs.  The added difficulty in
reaching the extreme energies in astrophysical sources further
justifies the search for alternative explanations. 
Future experiments will play a crucial  role in settling this long
standing mystery, through the determination of the cosmic ray flux,
and arrival direction distribution, and the great discriminator: the  
composition at extremely  high energies.
 
\section*{Acknowledgment} 
  
 This work was supported by NSF through grant 
AST-0071235  and DOE grant DE-FG0291  ER40606 at the University of 
Chicago and at Fermilab by DOE and NASA grant NAG 5-7092.


\begin{thebibliography}{9} 
 
\bibitem{data}  M. Takeda et al., Phys. Rev. Lett. 
81 (1998) 1163;  M. Takeda et al. preprint astro-ph/9902239 
(submitted to Astrophys. J.); N. Hayashida et al. Phys. Rev. Lett. 73 
(1994) 3491; D. J. Bird et al.  Astrophys J. 441 (1995) 
144;  Phys. Rev. Lett. 71 (1993) 3401;  
Astrophys. J. 424 (1994) 491; M. A. Lawrence,  R. J. O. Reid  
and A. A. Watson,  J. Phys. G. Nucl. Part. Phys.  17 (1991) 773; N. N. 
Efimov et al., Ref.  Proc. International Symposium on {\it Astrophysical  
Aspects of the Most Energetic Cosmic Rays}, eds. M. Nagano and F. 
Takahara (World Scientific, Singapore, 1991), p. 20; 
 D. Kieda et al., HiRes Collaboration,  Proceeds. 26th 
ICRC Salt Lake (1999); J. Linsley,  Phys. Rev. Lett. 10 (1963) 146.

\bibitem{GZK66}  K. Greisen, Phys. Rev. Lett.  16 (1966) 748; G. T.  
Zatsepin  and V. A. Kuzmin, Sov. Phys. JETP Lett. 4  (1966) 78. 
 
\bibitem{haya00} N. Hayashida et al. (2000) 
 
\bibitem{Olin00} A. Olinto, Phys.Rept. 333-334 (2000) 329. 
 
\bibitem{BS00}  P. Bhattacharjee and G. Sigl,    
Phys. Rept.  327 (2000) 109. 
 
\bibitem{BG88}  
V. Berezinsky and S. Grigorieva, Astron. Astroph.  199 (1988) 1. 
 
\bibitem{BG79}  
Berezinsky, V. S., and Grigorieva, S. I., {\it Proc. 16th. Int. Cosmic 
Ray Conf., Kyoto} 2 (1979) 81.  
 
\bibitem{MT99} G. A. Medina-Tanco,  Proceedings of 26th ICRC, Salt 
Lake City, Utah, vol 4, 346 (1999); Medina Tanco, G. A., Astrophys. J.  
510 (1999) 91.  
 
\bibitem{BW00} J. N.  Bahcall and E. Waxman, hep-ph/9912326v2 (2000). 
 
\bibitem{Ptu98} V.S. Ptuskin, S.I. Rogovaya, \& V.N. Zirakashvili,
Proceedings of 26th ICRC, Salt  Lake City, Utah, vol 4, 271 (1999).

\bibitem{branch98a} 
Branch, D.~1998, \araa, 36, 17 
 
\bibitem{davis82a} 
Davis, M., \& Huchra, J.~1982, \apj, 254, 437 
 
\bibitem{strauss95a}  
Strauss, M.~A., \& Willick, J.A.~1995, {Phys. Rep.}, {261}, 271 
  
\bibitem{huchra95a} 
Huchra, J.~P., Geller, M.~J., \& Corwin, Jr., H.~G.~1995, \apj, 70, 
687 
 
\bibitem{saunders00a} 
Saunders, W., Sutherland, W.~J., Maddox, S.~J., Keeble, O., Oliver, 
S.~J., Rowan-Robinson, M., McMahon, R.~G., Efstathiou, G., Tadros, H., 
White, S.~D.~M., Frenk, C.~S., Carraminana, A., Hawkins, 
M.~R.~S.~2000, submitted to \mnras, preprint (astro-ph/0001117) 
 
\bibitem{peebles93a} 
Peebles, P.~J.~E.~1993, Principles of Physical Cosmology (Princeton, 
NJ: Princeton University Press) 
 
\bibitem{hogg99a} 
Hogg, D.~W.~1999, astro-ph/9905116 
 
\bibitem{peebles80a} 
Peebles, P.~J.~E.~1980, The Large-Scale Structure of the 
Universe (Princeton, NJ: Princeton University Press) 
 
\bibitem{efstathiou88a} 
Efstathiou, G., Ellis, R.~S., \& Peterson, B.~S.~1988, \mnras, 232, 431 
 
\bibitem{sandage79a} 
Sandage, A., Tammann, G.~A., \& Yahil, A.~1979, \apj, 232, 352 
 
\bibitem{BBDGP90} V. S. Berezinsky, S.V. Bulanov, V. A. Dogiel,  
V. L. Ginzburg,  and V. S. Ptuskin,  Astrophysics of Cosmic Rays, 
(Amsterdam: North Holland, 1990). 

\bibitem{Acht98} Achterberg et al. 1998 
A. Achterberg, Y. A. Gallant, C. A. Norman, D. B. Melrose,
astro-ph/9907060

 
\bibitem{AGASA} 
Takeda, M., et al., {\it Phys. Rev. Lett.} {\bf 81}, 1163 (1998). 
 
\bibitem{AGASAnew} 
Hayashida, N., et al., Appendix to {\it Astrophys. J.} {\bf 522}, 225 (1999) 
(preprint astro-ph/0008102). 
 

\bibitem{BO99} P. Blasi and A. V. Olinto,  1999, Phys. Rev. D 59, 
023001. 
 
\bibitem{SLB99}  G. Sigl, M. Lemoine,  and P. Biermann,   
Astropart. Phys. 10 (1999) 141. 
 

\bibitem{stanev}
T. Stanev, R. Engel, A. Muecke, R. J. Protheroe, J. P. Rachen,
 astro-ph/0003484.

\bibitem{Auger} 
J. W. Cronin, Nucl. Phys. B. (Proc. Suppl.) 28B (1992) 213.
 
\bibitem{HiRes}  S. C. Corbat\'{o} et al., Nucl. Phys. B 
(Proc. Suppl.) 28B (1992) 36. 
 
\bibitem{TA}  M. Teshima et al., Nucl. Phys. B (Proc. Suppl.) 
28B (1992) 169.  
 
\bibitem{EUSO} see http://www.ifcai.pa.cnr.it/Ifcai/euso.html
 
\bibitem{OWL} R. E. Streitmatter, Proc. of {\it Workshop on Observing 
Giant Cosmic Ray Air Showers from $>10^{20}$ eV Particles from Space}, 
eds. J. F. Krizmanic, J. F. Ormes, and R. E. Streitmatter 
(AIP Conference Proceedings 433, 1997). 
 
\bibitem{BIGM} E. Waxman and  J. Miralda-Escude,  Astrophys. J. 472 
(1996) L89; G. A. Medina Tanco, E. M. de Gouveia Dal Pino, and 
J. E. Horvath, Astropart. Phys. 6 (1997) 337; G. Sigl, M. Lemoine, and A. 
V. Olinto, Phys. Rev. 56  (1997) 4470; M. Lemoine, G. Sigl, 
A. V. Olinto, and D. 
Schramm, Astropart. Phys.  486  (1997) L115; G. Sigl, M. Lemoine, 
and P. Biermann,   
Astropart. Phys. 10 (1999) 141; D. Ryu, H. Kang   and  P. L. Bierman,   
Astron. Astrophys. 335 (1998) 19. 
 
\bibitem{VLI}  L. Gonzalez-Mestres, Nucl. Phys. B 
(Proc. Suppl.) 48 (1996) 131; S. Coleman and S. L. Glashow, 
 Phys. Rev. D 59 (1999) 116008; 
 H. Sato and T. Tati, Prog. Theor. Phys. 47 (1972) 1788;  
 D. A. Kirzhnits and V. A. Chechin, Sov. J. Nucl. Phys. 
15 (1972) 585;  R. Aloisio, P. Blasi, P Ghia, 
and A. Grillo, preprint INFN/AE-99/24. 

\end{thebibliography}
\end{document}